# Philosophy of Mind and the Problem of Free Will in the Light of Quantum Mechanics.

*Henry P. Stapp*
*Lawrence Berkeley National Laboratory*
*University of California*
*Berkeley, California 94720*

## Abstract

Arguments pertaining to the mind-brain connection and to the physical effectiveness of our conscious choices have been presented in two recent books, one by John Searle, the other by Jaegwon Kim. These arguments are examined, and it is explained how the encountered difficulties arise from a defective understanding and application of a pertinent part of contemporary science, namely quantum mechanics. The principled quantum uncertainties entering at the microscopic levels of brain processing cannot be confined to the micro level, but percolate up to the macroscopic regime. To cope with the conflict between the resulting macroscopic indefiniteness and the definiteness of our conscious experiences, orthodox quantum mechanics introduces the idea of agent-generated probing actions, each of which specifies a definite set of alternative possible empirically/experientially distinguishable outcomes. Quantum theory then introduces the mathematical concept of randomness to describe the probabilities of the various alternative possible outcomes of the chosen probing action. But the agent-generated choice of which probing action to perform is not governed by any known law or rule, statistical or otherwise. This causal gap provides a logical opening, and indeed a logical need, for the entry into the dynamical structure of nature of a process that goes beyond the currently understood quantum mechanical statistical generalization of the deterministic laws of classical physics. The well-known quantum Zeno effect can then be exploited to provide a natural process that establishes a causal psychophysical link within the complex structure consisting of a stream of conscious experiences and certain macroscopic classical features of a quantum mechanically described brain. This naturally created causal link effectively allows consciously felt intentions to affect brain activity in a way that tends to produce the intended feedback. This quantum mechanism provides an eminently satisfactory alternative to the classical physics conclusion that the physical present is completely determined by the physical past, and hence provides a physics-based way out of the dilemma that Searle and Kim tried to resolve by philosophical analysis.

## 1. Introduction.

The central problem in philosophy of mind is the mind-body problem: the problem of reconciling our science-based understandings of the causal structure of the physically described world, including our bodies and brains, with the apparent capacity of our conscious thoughts and efforts to cause our bodies to move in consciously intended ways.

The contention of the present work is that the difficulties that philosophers of mind are encountering in coming to a satisfactory resolution of this problem arise from a faulty understanding and application of a relevant part of contemporary science, namely quantum mechanics. Philosophy of mind is a vast field, so to make my task manageable I shall limit my remarks to the opinions and arguments presented in two recent books, John Searle's *Freedom and Neurobiology: Reflections on Free Will, Language, and Political Power*[1], and Jaegwon Kim's *Physicalism, or Something Near Enough*[2].

This work was supported by the Director, Office of Science, Office of High Energy and Nuclear Physics, of the U.S. Department of Energy under contract DE-AC02-05CH11231

## 2. Searle's Approach.

John Searle begins Section 1 of his book with the assertion: "There is exactly one overriding question in contemporary philosophy…As a preliminary succinct formulation we could put it in these terms: How do we fit in?" He explains that: "We now have a reasonable well-established conception of the basic structure of the universe." "We understand that the universe consists entirely of particles (or whatever entities the ultimately true physics arrives at), and these exist in fields of force and are typically organized into systems." He observes that: "On our earth, carbon-based systems made of molecules that also contain a lot of hydrogen, nitrogen and oxygen have provided the substrate of human, animal and plant evolution", and says that: "These and other such facts about the basic structure of the universe, I will call, for short, the 'basic facts' . The most important sets of basic facts, for our present purposes, are given in the atomic theory of matter and the evolutionary theory of biology."

These statements identify the foundation and orientation of Searle's approach: We human beings are biological systems made of atoms and molecules, and our complete understanding of ourselves should therefore emerge from an analysis of our understandings of our biological structures, which rest in turn on the atomic theory of matter.

Searle notes that his approach rests also on an important difference between what is possible in philosophy today and what was feasible in the past. He notes that "For three centuries after Descartes, the epistemological questions, especially the skeptical questions, formed the center of philosophical interest. (p.26)" That quest can now be ended because "We simply know too much. We have a prodigious amount of knowledge that is known with objectivity, certainty, and universality. … They are known with certainty, in the sense that the evidence is now so great that it is irrational to doubt them."(p. 27)" Searle thus escapes the search-for-certainty dead end by accepting the above-mentioned 'basic facts'.

Searle observes that we have, however, in addition to the 'basic facts' also a conception of ourselves as conscious, intentionalistic, rational, …free will possessing agents," and he identifies the question to be addressed by his book--which is also the topic of this article--as: "How can we square this self-conception of ourselves as mindful, meaning-creating, free, rational, etc., agents with a universe that consists entirely of mindless, meaningless, unfree, nonrational, brute physical particles.(p.5)"

Two key problems facing this endeavor are "consciousness" and "free will". Searle claims to have solved the *philosophical* problem of consciousness by asserting that "Conscious states are entirely caused by neuronal processes in the brain, and are realized in the brain." The residual problems of consciousness are thereby relegated to neurobiology: "How exactly does the brain cause conscious experiences, and how are those experiences realized in the brain." (p.6)

## 3. Free Will.

The free-will problem is "How can there exist genuinely free actions in a world where all events, at least at the macro level, apparently have causally sufficient antecedent conditions? Every event at that level appears to be determined by causes that preceded it. Why should acts performed during apparent human consciousness of freedom be an exception? It is true that there is an indeterminacy in nature at the quantum level, but that indeterminacy is pure randomness and randomness is not by itself sufficient to give free will." (p. 10-11)

Searle admits that for the problem of free will "we are nowhere remotely near having a solution." (p.11)

Why is solving the problem of free will so important?

The useful practical purpose of philosophy is to arrive at a coherent understanding of how we fit in, in order that we may conduct our lives in accordance with principles not beset with contradictions. Searle makes a convincing case (p.11) that we must, in order to function rationally in this world, believe that we are sometimes free to choose our actions. To deny this would create a self-contradiction. But without resolving the problem of free will, philosophy loses its rational coherence, and men will turn to other sources for the foundations of their beliefs.

The importance of arriving at a solution of the free-will problem is highlighted also by recent controlled studies[3] that show that experimental subjects conditioned by arguments that promote the thesis that we have no free will, that free will is an illusion, that mind is epiphenomenal, are more likely to cheat and lie than subjects conditioned by arguments defending the thesis that the freedom that we feel is bona fide. Hence, again, achieving a solution of the free-will problem has important consequences in our lives.. This motivates our taking a closer look at Searle's arguments and the difficulties that they create for solving the free-will problem.

Searle, following his neurobiological approach, must explain how free will can be converted to a problem in neurobiology. He considers two hypotheses (p.61-73):

Hypothesis 1: The neurobiological state of the brain is causally sufficient to determine the behavior of the brain, hence the body. In this case, the *feeling of freedom to choose* some of our actions is an illusion! Consciousness lacks causal efficacy. It is purely epiphenomenal. Searle emphasizes that this idea---that nature has provided us with this fantastic feature, consciousness, that seems to play an essential role in the successful conduct of our lives, but that actually does nothing---is "unattractive" (p.70). He emphasizes the great biological cost of producing the machinery needed to create consciousness. The suggestion that the output of this costly biological process has no physical effect is hard to square with evolutionary theory.

Hypothesis 2: The neurobiological state of the brain is causally *in*sufficient to determine the behavior of the brain, and this causal gap allows our conscious choices to influence our conduct in the way that they seem to do, namely on the basis of *choices based on reasons*. He argues that "reasons" can fulfill the role of sufficient conditions only by way of influencing our deliberating, choosing, and physically efficacious conscious "selves".

The demand that neurobiological state of the brain is causally insufficient to determine the ongoing behavior of the brain entails the failure of one of the chief properties of classical physical theories, the causal closure of the physically described aspects of nature. Searle says:

"It is tempting, indeed irresistible, to think that the explanation of the conscious experience of free will must be a manifestation of quantum indeterminism at the level of conscious, rational decision making. Previously I never could see the point of introducing quantum mechanics into the discussion of consciousness. But here is at least a strict argument requiring the introduction of quantum indeterminism.

Premise 1. All indeterminism in nature is quantum indeterminism.

Premise 2. Consciousness is a feature of nature that manifests indeterminism.

Conclusion: Consciousness manifests quantum indeterminism."

"….This [conclusion] is important for contemporary research. The standard lines of research…make no appeal to quantum mechanics. If Hypothesis 2 is true these cannot succeed, at least not for volitional consciousness." (p.75)

He goes on: "If quantum indeterminism amounts to randomness then quantum indeterminism by itself seems useless in explaining the problem of free will because free actions are not random." The point here is that we need an explanation not only of the *failure* of physical determinism, but an explanation also of the filling of that causal gap by our "free" choices based on *reasons*.

Summarizing, he says: "Once we sorted out the issues we found two possibilities, Hypothesis 1 and Hypothesis 2. Neither is very appealing. If we had to bet, the odds would surely favor Hypothesis 1, because it is simpler and fits with our overall view of biology. But it gives results that are literally incredible." (One cannot literally believe that oneself cannot make choices.) But "Hypothesis 2 is a mess, because it gives us three mysteries for one. We thought free will was a mystery, but consciousness and quantum mechanics were two separate and distinct mysteries. Now we have the result that in order to solve the first we have to solve the second and invoke one of the most mysterious aspects of the third to solve the first two."

**4. The Three Mysteries.**

But who could think that these three "mysteries" were separate and distinct?

With regard to the connection between free will and mind, William James asserted, near the beginning of *The Principles of Psychology*[4],

> "*The pursuance of future ends and the choice of means for their attainment are thus the mark and criterion of the presence of mentality* in a phenomenon". (James, p.8)

> "*No actions but such as are done for an end, and show a choice of means, can be called indubitable expressions of Mind*". (James, p.10)

Thus, for James, mind is fundamentally tied to *the choice of a means to an end.* On the other hand, the solution that Searle offered long ago to the mind-brain problem did not touch on free will. It said simply: "Conscious states are entirely caused by neuronal processes in the brain, and are realized in the brain." He notes, as mentioned above, that this *philosophical* solution relegates to neurobiology the residual questions: "How exactly does the brain cause conscious experiences?" "How are those experiences realized in the brain." And he admits that contemporary mainstream neurobiology is nowhere near solving these residual questions. Indeed, insofar as neurobiology bases itself purely on classical mechanics, it lacks any logical or theoretical basis to link the empirically observed correlations between conscious experiences and brain behavior to any notion of how this classically conceived physically described brain could *cause to occur events having the knowingness and feelingness that characterize our conscious experiences*. There is nothing in the classical conception of physically described matter that could *cause* (even) a complex classically conceived high-level systems property to embellish itself, or endow itself, with *an experience of knowing or feeling*. Such a causal capacity is not in the inventory of properties assigned to physically described systems by classical physics. The physically described aspects of systems, as conceived of in classical physics---*unlike the physically described aspects of systems as conceived of in quantum mechanics*---have been stripped of any necessary causal connection to knowings or feelings. The physical aspects are both causally and conceptually complete. Thus, insofar as the neurobiology that Searle contemplates is based fundamentally on the classical physics of the seventeenth through nineteenth centuries, it is not true that the 'basic facts' entail a neurobiological solution of the mind-body problem of the kind that Searle asserts. There is a logical gap! A true 'basic fact' is that this classical conception of the physical is inadequate to explain the atomic properties upon which actual neurobiological structures are based. Because Searle has not incorporated the logical and causal structure of quantum mechanics into his conception of neurobiology, his claim to have solved even the purely philosophical part of the mind-body problem is not rationally justified: there is no rational reason to believe that a solution along classical lines is possible in a fundamentally non-classical universe, particularly since the orthodox quantum successor to classical physics involves the *necessary* introduction, into the basic dynamics, of actions by agents; actions that are not specified by the micro-physical laws, but that, within the theory, arise from *free choices of means to attain intended ends*.

This flaw is implicit in Searle's open-ended introductory proviso, i.e. "(or whatever entities the ultimately true physics arrives at)." His arguments tacitly assume that these entities will be like "quarks" or other "mindless" entities, not like the mindful elements of our streams of consciousness. Yet

Searle's 'basic facts' include atomic theory, which was radically transformed during the twentieth century. Searle uses the new theory, quantum theory, in his analysis of free will. However, the opening words of Bohr's 1934 book *Atomic Theory and the Description of Nature*[5] are: "The task of science is both to extend the range of our experience and reduce it to order." This idea is restated many times in many ways, for example as: "In our description of nature the purpose is not to disclose the real essence of phenomena but only to track down as far as possible relations between the multifold aspects of our experience. (p.18)" Werner Heisenberg's famous expression of this point was:

> "The conception of the objective reality of the elementary particles has thus evaporated not into the cloud of some obscure new reality concept but into the transparent clarity of a mathematics that represents no longer the behavior of the particles but rather our knowledge of this behavior" (Heisenberg[6] , p.100)

These statements emphasize that the basic ontological realities of quantum theory are not physical particles, but rather increments in knowledge. They are conscious experiences occurring in streams of conscious experiences. The "physical description" of earlier (classical) physical theories is *transformed* in quantum mechanics to a mathematical structure that represents not  material particles but rather "potentia" (objective tendencies) for new knowledge-increasing events to occur in our streams of consciousness. Each such event is accompanied by a change in the mathematically described "potentia" for future events. This change renders the potentialities for future experiences consistent with the increased knowledge. The theory is therefore useful and testable because it directly predicts *relationships between our experiences---between our conscious acts of knowing*.

Searle introduces, in connection with his analysis of free will, the indeterminacy aspect of quantum mechanics but not the other profoundly relevant features just mentioned. On page 11 he says "It is true that there is an indeterminacy in nature at the quantum level, but that indeterminacy is pure randomness and randomness is not by itself sufficient to give free will."

While it is absolutely true that randomness is different from, and insufficient for, free will, which involves reason-based choices of means to attain intended ends, it is *absolutely untrue* that quantum indeterminism is pure randomness.  Bohr says:

> To my mind there is no alternative than to admit that, in this field of experience, we are dealing with individual phenomena and that our possibilities of handling the measuring instruments allow us only to make a choice between the different complementary types of phenomena that we want to study. (Bohr[7], p.51)

> The freedom of experimentation, … is fully retained and corresponds to the free choice of experimental arrangement for which the mathematical structure of the quantum mechanical formalism offers the appropriate latitude[6].  (Bohr[7], p. 71)

This "appropriate latitude" "offered by the mathematical structure of the quantum mechanical formalism" is a key aspect of the indeterminism of quantum mechanics, but it is quite different from another aspect which is the "randomness". The quantum mechanical approach rests on a postulated connection between two aspects of the scientific description of phenomena. One aspect is described in terms of the mathematical structure of the quantum mechanical formalism. The other aspect is described in terms of appearances:

> …we must recognize above all that, even when the phenomena transcend the scope of classical physics, the account of the experimental arrangements must be given in plain language, suitably supplemented by technical physical terminology. This is a simple logical demand, since the very word "experiment" refers to a situation where we can tell others what we have done and what we have learned. (Bohr[7], p.72)

Von Neumann[8] formulated quantum mechanics in a mathematically and logically rigorous way. He gave the name "process 1" to the *physically described counterpart* of the "free choice of experimental arrangement" that is described in terms of appearances/experiences.  A key feature of this choice, and hence of its process 1 physical counterpart, is that it is "free" in the specific sense that the quantum laws and rules place no conditions, *statistical or otherwise*, upon it! This choice is free of any known theoretical constraint! It is indeterminate. The 'random' aspect comes in, logically, only *after* this process 1 physical action. The process 1 action specifies some particular *partition* of the prior physical state into a countable set of distinct possibilities. Orthodox quantum theory then asserts that some single one of the *specified-by-process-1 distinct possibilities* will occur, randomly. The "randomness" condition asserts that these *occurrences of outcomes* will be in accord with statistical weights that are specified by the quantum mechanical formalism. But the preceding *partitioning of the prior collection of possibilities into a (countable) set of distinct possibilities is an indeterminate processes that according to orthodox quantum mechanics is not random, but is treated, rather, as a reason-based choice of means to an end!* The experimenter chooses between this set-up or that set-up on the basis of reasons! More generally, a person's most consciously made choices appear to arise from reasons, and feelings. And within quantum mechanics there is a *logical need* for physically effective choices that are not determined, even statistically, by the known laws of quantum mechanics. There is, therefore, a rational opening for the causal roots of the causally effective choices mandated by quantum theory to lie, in part, in the realm of our streams of consciousness, and hence for the principle of the causal closure of the physical to fail.

Von Neumann's work allows our bodies and brains to be described in terms of the quantum mechanical formalism. This makes the two disparate descriptions---perceptual-intentional and quantum-mathematical---that occur in quantum mechanics identifiable with the two disparate descriptions occurring in the mind-body problem.

What these features of quantum mechanics imply is that Searle's basic idea that the "basic facts" entail "a universe that consists entirely of mindless, meaningless, unfree, nonrational, brute physical particles" is grossly at odds with a universe containing ourselves in the way specified by quantum mechanics. The reduction to "brute physical particles" is a feature that emerges only in the classical approximation.

Searle concludes (p.71) that "It seems to me that there are three conditions, in ascending order of difficulty, and an account of brain functioning in accord with Hypothesis 2 would have to explain how the brain meets these conditions."

1. "Consciousness, as caused by neuronal processes and realized in neuronal systems, functions causally in moving the body."

2. "The brain causes and sustains the existence of a conscious self that is able to make rational decisions and carry them out in actions."

3. "The brain is such that the conscious self is able to make and carry out decisions in the gap, where neither decision nor action is determined in advance by causally sufficient conditions, yet both are rationally explained by the reasons the agent is acting on."

As regards condition 1, Searle claims that he has already explained how this is possible, by analogy with the Roger Sperry's example of how the "solidity" (a high-level property) allows the motion of the whole wheel to cause its molecules to move in a coordinated way, controlled from top-down by high-level collective properties. But he had already noted (p. 64) that: "any analogy goes only so far. The analogy, solidity is to molecular behavior as consciousness is to neuronal behavior, is inadequate at, at least, two points. First, we take the wheel to be entirely deterministic, and ... second, the solidity of the wheel ontologically reducible to the behavior of the molecules", whereas neither of these conditions carry over to the Hypothesis 2 case. Hence the basis of his earlier claim to have solved the consciousness (mind-body) problem disintegrates in the Hypothesis 2 case. There is no need, on the basis of the *true* 'basic facts', for consciousness to be (fully) caused by the brain, as the brain is described in quantum mechanics.

As regards Searle's condition 2, the quantum ontological foundation of a person's stream of conscious experiences is no longer solely a classically conceived brain. The quantum mechanically conceived brain specifies only the potentialities/probabilities for certain psychophysical events to occur *under the condition that certain associated described process 1 choices have previously been made*. These dual-aspect psychophysical events are the basic entities. The psychological description and the physical description specify two aspects (sides) of a single event-type entity. The conscious self is a stream of conscious events. These events are the psychologically described aspects of a sequence of psychophysical events whose physical aspects are a sequence of physical events in a single brain. Mental process is to be understood in terms of this richer dualistic ontological base, rather than the impoverished purely physical part that survives contraction to the classical approximation. So it is not completely evident that the brain "causes" the conscious self, as Searle avers, as contrasted to the possibility that the detailed structure of the evolving brain comes to be what it is at any moment in time by virtue of a process that is more of a collaboration between mind and brain dependent in part upon the physical effects of the conscious-agent-generated process 1 physical process than exclusively on a one-way bottom up causal process.

The way that condition 3 is satisfied in quantum mechanics is described in sections 7 and 8, and relies heavily on the quantum Zeno effect.

I turn next to a discussion of Kim's arguments.

## 5. Physicalism.

The widely held philosophical position called "physicalism" has been described and defended in a recent book by Jaegwon Kim[2]. The physicalist claims that the world is basically purely physical. However, "physical" is interpreted in a way predicated, in effect, upon certain properties of classical physics that are contravened by the precepts of orthodox quantum physics. Kim's arguments reveal two horns of a dilemma that the physicalist is forced to face as a consequence of accepting the classical notion of "physical". Kim admits that neither of the two options, "epiphenomenalism" or "reduction", is very palatable, but he finds a compromise that he deems acceptable.

The central aim of the present article is to show that the physicalist's dilemma dissolves when one shifts from the classical notion of the physical to its quantum mechanical successor. Understanding this shift involves distinguishing the quantum conception of the mind-brain from the shadow of itself that survives reduction to the classical approximation.

To make clear the essential features of the quantum mechanical conception of the mind-brain connection, I shall describe here a model that is a specific realization of a theory I have described in more general terms before[9-12]. Being specific reduces generality, but having a concrete model can be helpful in revealing the general lay of the land. Also, the specific features added here resolve in a natural way the puzzle of how our descriptions of our observations can be couched in the language of classical physics when our brains are operating, fundamentally, in accordance with the principles of quantum theory. The specific model also shows how the thoroughly quantum mechanical (quantum Zeno) effect, which underlies the power of a person's conscious thoughts to influence in intended ways the physically described processes occurring in that person's brain, is not appreciably disrupted either by "environmental decoherence" effects or by thermal effects arising from the "hotness" of the brain.

In order to communicate to the broad spectrum of scientists and philosophers interested in the connection between mind and brain, and in the issue of free will, I will review in the following section the historical and conceptual background of the needed quantum mechanical ideas, and then describe an approach to the mind-body problem that is based fundamentally on quantum theory, but that adds several specific extra ideas about the form of the mind-brain connection.

## 6. Quantum Mechanics and Physicalism.

Rather than just plunging ahead and using the concepts and equations of quantum mechanics, and thereby making this work unintelligible to many people that I want to reach, I am going to provide first an historical and conceptual review of the extremely profound changes in the philosophical and technical foundations that were wrought by the transition from classical physics to quantum physics. One key technical change was the shift from the *numbers* used in classical mechanics to describe properties of physical systems to the associated *operators* or *matrices* used to describe related *actions*. This technical shift emerged, unsought, from a seismic conceptual shift. Following the path blazed by Einstein's success in creating special relativity, Heisenberg changed course. Faced with a quarter century of failures to construct a successful atomic theory based upon the notion of some presumed-to-exist space-time structure of the atom, Heisenberg attempted to build a theory based upon our observations and measurements, rather than upon conjectured microscopic space-time structures that could be *postulated* to exist, but that were never directly observed or measured. This shift in orientation led to grave issues concerning exactly what constituted an "observation" or "measurement". Those issues were resolved by shifting from an ontological perspective---which tries to describe what really exists objectively "out there"--- to a practical or pragmatic perspective, which regards a physical theory as a useful collective conceptual human endeavor that aims to provide us with reliable expectations about our future experiences, for each of the alternative possible courses of action between which we are (seemingly) free to choose. As a collective endeavor, and in that sense as an objective theory, quantum mechanics is built on *descriptions* that allow us to communicate to others what we have done and what we have learned. Heisenberg strongly emphasized that this change in perspective converts the quantum mechanics, in a very real sense, into a theory about "our knowledge": the relationships between experiential elements in our streams of consciousness become the core realities of a conceptual construction that aims to allow us to form, on the basis of what we already know, useful expectations about our future experiences, under the various alternative possible conditions between which we seem able to freely choose.

The paradoxical aspect of claiming the "physical state of a system" to be a representation of "our knowledge" is starkly exhibited by "Schroedinger's cat", whose quantum state is, according to this pragmatic approach, not determined until someone looks. Bohr escapes this dilemma by saying that principles of his (Copenhagen) approach are insufficient to cover biological systems. But that limitation leaves quantum mechanics fundamentally incomplete, and, in particular, inapplicable to the physical processes occurring in our brains.

In an effort to do better, von Neumann[8] showed how to preserve the rules and precepts of quantum mechanics all the way up to the mind-brain interface, preserving the basic character of quantum mechanics as a theory that aims to provide reliable expectations about future experiences on the basis of present knowledge. Von Neumann's work brings into sharp focus the central problem of interest here, which is the connection between the properties specified in the quantum mechanical description of a person's brain and the experiential realities that populate that person's stream of consciousness. Bohr was undoubtedly right in saying that the Copenhagen precepts would be insufficient to cover this case. Additional ideas are needed, and the purpose of this article is to provide them.

The switch from classical mechanics to quantum mechanics preserves the idea that a physical system has a physically describable state. But the character of that state is changed drastically. Previously the physical state was conceived to have a well defined meaning independently of any "observation". Now the physically described state has essentially the character of a "potentia" (an "objective tendency") for the occurrence of each one of a continuum of alternative possible "events". Each of these alternative possible events has both an experientially described aspect and also a physically described aspect: each possible "event" is a psycho-physical happening. The experientially described aspect of an event is an element in a person's stream of consciousness, and the physically described aspect is a *reduction* of the set of objective tendencies represented by the prior state of that person's body-brain to the *part* of that prior state that is compatible with the increased knowledge supplied by the new element in that person's stream of consciousness. Thus the changing psychologically described state of that person's knowledge is correlated to the changing physically described state of the person's body-brain, and the changing physically described state entails, via the fundamental quantum probability formula, a changing set of weighted possibilities for future psychophysical events.

The practical usefulness of quantum theory flows from this lawful connection between a person's increasing knowledge and the changing physical state of his body-brain. The latter is linked to the surrounding physical world by the dynamical laws of quantum physics. This linkage allows a person to "observe" the world about him by means of the lawful relationship between the events in his stream of conscious experiences and the changing state of his body-brain.

It is worth noting that the physically described aspect of the theory has lost its character of being a "substance", both in the philosophical sense that it is no longer *self-sufficient,* being intrinsically and dynamically linked to the mental, and also in the colloquial sense of no longer being *material*. It is *stripped of materiality* by its character of being merely a collection of potentialities or possibilities for future events. This shift in its basic character renders the physical aspect somewhat idea-like, even though it is conceived to represent objectively real tendencies.

The key "utility" property of the theory---namely the property of being useful---makes no sense, of course, unless we have, in some sense, some freedom to choose. An examination of the structure of quantum mechanics reveals that the theory has both a logical place for, and a logical need for, choices that are made in practice by the human actor/observers, but that are *not determined by the quantum physical state of the entire world, or by any part of it*. Bohr calls this choice "the free choice of experimental arrangement for which the quantum mechanical formalism offers the appropriate latitude." (Bohr[7], p.73). This "free" choice plays a fundamental role in von Neumann's rigorous formulation of quantum mechanics, and he gives the physical aspect of this probing action the name "process 1" (von Neumann, p. 351, 418, 421). This process 1 action is not necessarily determined, even statistically, by the physically described aspects of the theory.

The fact that this choice made by the human observer/agent is not necessarily determined by the physical state of the universe means that *the principle of the causal closure of the physical domain is not necessarily maintained in contemporary basic physical theory*. It means also that Kim's formulation of *mind-body supervenience is not entailed by contemporary physical theory*. That formulation asserts that "what happens in our mental life is wholly dependent on, and determined by, what happens with our bodily processes ." (p. 14) Kim indicates that supervenience is a common element of all *physicalist theories*. But since this supervenience property is not required by basic (i.e., quantum) physics, the easy first step out of the difficulties that have been plaguing physicalists for half a century, and that continue to do so, is simply to recognize that the precepts of classical physics, which are the scientific source of the notions of the causal closure of the physical, and also of this idea of supervenience, do not hold in real brains, whose activities are influenced heavily by quantum processes that require (process 1) physical inputs that are not necessarily *wholly* determined by what happens with our bodily processes.

Before turning to the details of the quantum mechanical treatment of the relationship between mind and brain I shall make a few comments on Kim's attempted resolution of the difficulties confronting the classical physicalist approach. The essential problem is the mind-body problem. Kim divides this problem into two parts, the problem of mental causation and the problem of consciousness. The problem of mental causation is: "How can the mind exert its causal powers in a world that is fundamentally physical?" (Kim, p.7) The problem of consciousness is: "How can there be such a thing as consciousness in a physical world, a world consisting ultimately of nothing but bits of matter distributed over space-time in accordance with the laws of physics." (Kim, p. 7)

From a modern physics perspective the way to resolve these problems is immediately obvious: Simply recognize that the assumption that the laws of physics pertain to "bits of matter distributed over space-time in accordance with the laws of physics" is false. Indeed, that idea has, for most of the twentieth century, been asserted by orthodox physicists to be false, along with the assumption that the world is physical in the classical sense. Quantum mechanics builds upon the undeniable real existence of our streams of conscious experiences, and provides also, as we shall see, a natural explanation of their causal power to influence physical properties. Thus the difficulties that have beset physicalists for decades, and have led to incessant controversies and reformulations, stem, according to the perspective achieved by twentieth century physics, directly from the fact that the physicalist assumptions not only do not follow from basic precepts of physics, but, instead, directly contradict them. The premises of classical physicalists have been incredibly out of step with the physics of their day.

Kim's "physicalist" solution to the problem of the connection between mind and brain is essentially to separate a mental reality such as a "pain", by dividing  "being in pain" into "the conscious experience of being in pain" and the "state of being in pain", and allowing the latter to be characterized as being caused by certain physically described causes and as causing some physically described effects/behaviors. The second part can be physical, and hence mind-body "physicalism" is achieved, except for the fact that the first part, "the conscious experience of being in pain", is non-physical and epiphenomenal.  Kim claims that this is the best that can be done by way of saving mind-body physicalism, but that this is "near enough". However, the epiphenomenal character of our streams of conscious experiences within classical-physics-based ontologies has always been the central problem, and Kim's physicalist "near enough" solution does not really solve it.

Kim tries in his chapter 3 to squash the notion that the difficulties with physicalism can be avoided by accepting some form of dualism. But the dualism that he considers is a Cartesian dualism populated on the mind side with mysterious disconnected "souls" whose "essential nature is that they are wholly outside the spatial order and lack all spatial properties". (p. 87). However, the experientially described mental entities that occur in pragmatic quantum theory are the basic realities of science. They are the ideas that we are able to communicate to others pertaining to what we have done and what we have learned. These descriptions are essentially descriptions of (parts of) the accessible contents of the streams of consciousness of real living observer-agents. In ontologically construed orthodox quantum mechanics these descriptions are descriptions of mental idea-like aspects of *real actual events,* each of which has also a physically described aspect that imposes in the spatio-temporally-based realm of potentialities for future psycho-physical events the  conditions entailed by the increment in knowledge that constitutes its mental aspect. Criticizing dualism in the soul-based (essentially disconnected) form advanced by Descartes during the seventeenth century instead of in the dual-aspect reduction-event form implicated by contemporary science is an indication that philosophers of mind have isolated themselves in a hermetically sealed world, created by considering only what other philosophers of mind have said, or are saying, with no opening to the breezes that bring word of the highly pertinent revolutionary change in the role of our conscious minds that had occurred in basic science during the 1920s.

Kim's chapter 3 is supposed to rule out dualism. But the dualism that he mainly addresses is a stark Cartesian (substance) dualism involving "souls" existing "outside physical space". He says "My target will be the interactionist dualism of Descartes." But Quantum mechanics involves a particular kind of dualism: it is a dual-aspect theory. In footnote 3 on page 71 Kim suggests that "dual-aspect" theories are "only variants of property dualism." He says later that "What has become increasingly evident over the past thirty years is that mental causation poses insuperable difficulties for all forms of mind-body dualism---for property dualism no less than substance dualism. …but I believe that if we have learned anything from the three decades of debate, it is that unless we bring the supposed mental causes fully into the physical world there is no hope of vindicating their status as causes, and that the reality of mental causation requires reduction of mentality to physical processes, or of minds to brains." (p.156). He gives on the preceding page a supposed way of "generating the problem of mental causation for property dualism" without assuming "the causal closure of the physical". But his argument includes an assumption "Given that your finger twitching, a physical event, has a full physical cause." This assumption is indeed less than an assumption of full causal closure of the physical. But in the quantum mechanical explanation of the way that mind causes bodily action the twitching does not have "a fully physical cause." According to quantum mechanics there needs to be a process 1 action mediating the connection between the psychologically described cause---a pain in this case---and any physical action caused by the pain. But the process 1 action has no known or necessary fully physical cause. A quantum mechanical account of how consciousness, per se, becomes causally effective is described in sections 7 and 8. It does not "bring the …mental causes fully into the physical world" but rather brings only the *effects* of the mental causes into the physical world.

Searle also has a problem with dualism. He says: "I am rejecting …any form of dualism.   Dualism is usually defined as the view that we live in two distinct realms, …the mental and the physical. The problem with dualism is that it amounts to giving up on the central enterprise of philosophy. …It might turn out, for example, that after our bodies are destroyed, our souls or conscious states will float about in a disembodied fashion. But it would be giving up on the philosophical (not to mention scientific) enterprise of trying to explain what we know to be real phenomena if we say that they defy explanation because they inhabit a separate realm." (p. 19).

But is dualism (in any form) usually the view that we live in two distinct realms?

The Oxford Companion to Philosophy (T. Honderich, ed.) defines:

Dualism: The theory that mind and matter are two distinct things. (p. 206)

The Blackwell "A Companion to Philosophy of Mind" (S. Guttenplan ed,) says:

…the dualist answer is that each person's mind is at least not identical with his body, so these are two different things. (p. 256)

Quantum mechanics allows the physical and mental aspects of a psychophysical event to be non-identical, while not saying that real phenomena defy explanation because they inhabit a separate realm. Rather it *explains* how conscious mental intentions can, *by virtue of the quantum mechanical laws themselves*, have the intended physically described effects.

Philosophers of mind appear to have arrived, today, at less-than-satisfactory solutions to the mind-brain and free will problems, and the difficulties seem, *at least prima facie*, very closely connected to their acceptance of a known-to-be-false understanding of the basic nature of the physical world, and of the causal role of our conscious thoughts within it.

In the following two sections I shall explain how these difficulties can be resolved by accepting an ontological construal of the essentially orthodox (von Neumann/Heisenberg) quantum mechanical understanding of the mind-body connection. By "orthodox" I mean an understanding that accepts the existence of reduction events that coordinate increments of knowledge to reductions of the physically described potentialities for future events.

7. **Quantum Mechanics: The Rules of the Game.**

7.1 The two basic formulas.

Quantum mechanics is a conceptual structure erected upon a certain kind of mathematical description of physical states, and a certain kind of phenomenal description of conscious experiences, and upon two basic formulas that connect these two kinds of descriptions. The physically described state of the universe, or of any physically describable subsystem, is represented by a mathematical structure called a density matrix or probability operator. It is usually represented in the theory by the symbol $\rho$. The first basic formula specifies the action upon the prior physical state $\rho$ by a *process 1 physical action*. This physically described action is tied, both conceptually and causally, to an associated "free choice of probing action described in everyday language, refined by the concepts of classical physical theory". An elementary process 1 action partitions the prior state $\rho$ into two distinct non-interfering parts.

$\rho \rightarrow P \rho P + P' \rho P'$     (P'= 1-P)

The first part, $P \rho P$, is associated with the occurrence/appearance of a pre-specified, possible, perceptually identifiable outcome 'Yes'. The other part is associated with a failure of that 'Yes' outcome to occur/appear.

The symbol $P$ represents an operator that satisfies $PP=P$. Such an operator is called a *projection operator*.

The quantum game is like "twenty questions": the observer-agent "freely poses" a question with an experientially identifiable answer 'Yes'. The physical process 1 probing action corresponding to this question is represented in the mathematical formalism by a projection operator $P$. Nature then returns an answer 'Yes' or 'No'. The probability that the experienced answer is 'Yes' is given by the basic probability formula of quantum mechanics:

$< P > = $ *Trace* $P \rho P$ */Trace* $\rho$.

The "Trace" operation acting upon an operator X, which is conceived of as an action that acts on whatever stands on its right, is the instruction: "Let X act back around upon itself!" The result is always a number. [For a detailed explanation of the mathematical meaning and workings of the above basic formulas see: http://arXiv.org/abs/0803.1625]

These two formulas constitute the foundation of the quantum mechanical rules for predicting certain statistical connections between: (1), the aspects of our conscious experiences that are described in the language that we use to describe to ourselves and to others the perceptual contents of our streams of conscious experiences; and (2), the aspects that are described in the mathematical formalism of quantum physics.

## 7.2 Classical Description.

"…we must recognize above all that, even when phenomena transcend the scope of classical physical theories, the account of the experimental arrangement and the recording of observations must be given in plain language, suitably supplemented by technical physical terminology. This is a clear logical demand, since the very word "experiment" refers to a situation where we can tell others what we have done and what we have learned." (Bohr[7], p. 72)

"…it is imperative to realize that in every account of physical experience one must describe both experimental conditions and observations by the same means of communication as the one used in classical physics." (Bohr[7], p. 88)

This demand that we *must use* the known-to-be-fundamentally-false concepts of classical physical theories as a fundamental part of quantum mechanics has often been cited as the logical incongruity that lies at the root of the difficulties in arriving at a rationally coherent understanding of quantum mechanics: i.e., of an understanding that goes beyond merely understanding how to use it in practice. So I will consider next the problem of reconciling the quantum and classical concepts, within the context of a quantum theory of the mind-brain connection.

## 7.3 Quasi-Classical States of the Electromagnetic Field

There is one part of quantum theory in which a particularly tight and beautiful connection is maintained between classical mechanics and quantum mechanics. This is the simple harmonic operator (SHO). With a proper choice of units the energy (or Hamiltonian) of the system has the simple quadratic form E = H= ½ ($p^2$ + $q^2$), where q and p are the coordinate and momentum *variables* in the classical case, and are the corresponding *operators* in the quantum case. In the classical case the trajectory of the "particle" is a circle in q-p space of radius r = $(2E)^{1/2}$. The angular velocity is constant and independent of E, and in these special units is ω = 1: one radian per unit of time. The lowest-energy classical state is represented by a point at rest at the "origin"  q = p = 0.

The lowest-energy quantum state ρ is the projection operator P corresponding to a Gaussian wave function that in coordinate space is
ψ(q) = C exp(— (½)$q^2$) and in momentum space is
ψ(p) = C exp(— (½ )$p^2$), where C is $2^{1/4}$.  If this ground state is shifted in q-p space by a displacement (Q, P) one obtains  a state---i.e., a projection operator---$P_{[Q,P]}$, which has the following important property: if one allows this quantum state to evolve in accordance with the quantum mechanical equations of motion then it will evolve into the trjectory of states $P_{[Q(t),P(t)]}$, where the (center) point (Q(t), P(t)) moves on a circular trajectory that is identical to the one followed by the classical point particle.

If one puts a macroscopic amount of energy E into this quantum state then it becomes "essentially the same as" the corresponding classical state. Thus if the energy E in this one degree of freedom is the energy per degree of freedom at body temperature then the quantum state, instead of being confined to an *exact point* (Q(t), P(t)) lying on a circle of (huge) radius r = $10^7$ in q-p space, will be effectively confined, due to the Gaussian fall-off of the wave functions, to a disc of *unit radius* centered at that point (Q(t), P(t)). Given two such states, $P_{[Q,.P]}$, and $P_{[Q',P']}$,  their overlap, defined by the Trace of the product of these two projection operators, is exp(— (½ )$d^2$), where d is the distance between their center points. On this $10^7$ scale the unit size of the quantum state becomes effectively zero. And if the energy of this classical SHO state is large on the thermal scale then its motion, as defined by the time evolution of the projection operator $P_{[P(t),Q(t)]}$, will be virtually independent of the effects of both environmental decoherence, which arises from subtle quantum-phase effects, and thermal noise, for reasons essentially the same as the reasons for the negligibility these effects on the classically describable motion of the pendulum on a grandfather clock.

Notice that the quantum state, $P_{[Q,.P]}$,  is completely specified by the corresponding classical state (Q, P): the quantum mechanical spread away from this point is not only very tiny on the classical scale; it is also completely fixed: the width of the Gaussian wave packet associated with our Hamiltonian is fixed, and independent of both the energy and phase of the SHO.

We are interested here in brain dynamics. Everyone admits that at the most basic dynamical level the brain must be treated as a quantum system: the classical laws fail at the atomic level. This dynamics rests upon myriads of microscopic processes, including flows of ions into nerve terminals. These atomic-scale processes must in principle be treated quantum mechanically. But the effect of accepting the quantum description at the microscopic level is to inject quantum uncertainties/indeterminacies at this level. Yet introducing even small uncertainties/indeterminacies at microscopic levels into these nonlinear systems possessing lots of releasable stored chemical energy has a strong tendency---the butterfly effect---to produce very large macroscopic effects later on. Massive parallel processing at various stages may have a tendency to reduce these indeterminacies, but it is pure wishful thinking to believe that these indeterminacies can be completely eliminated in all cases, thereby producing brains that are completely deterministic at the macroscopic level. *Some* of the microscopic quantum indeterminacy *must* at least occasionally make its way up to the macroscopic level.

According to the precepts of orthodox quantum mechanics, these macroscopic quantum uncertainties are resolved by means of process 1 interventions, *whose forms are not specified by the quantum state of the universe, or any part thereof*. In actual practice, what happens is determined by conscious choices "for which the quantum mechanical formalism offers the appropriate latitude". No way has yet been discovered by quantum theorists to circumvent this need for some sort of intervention that is not determined by the orthodox physical laws of quantum physics. In particular, environmental decoherence effects certainly do not, by themselves, resolve this problem of reconciling the quantum indeterminacy, which irrepressibly bubbles up from the microscopic levels of brain dynamics, with the essentially classical character of our descriptions of our experiences of "what we have done and what we have learned".

The huge importance of the existence and properties of the quasi-classical quantum states of SHOs is this: If the projection operators P associated with our experiences are projection operators of the kind that instantiate these quasi-classical states then we can rationally reconcile the demand that the dynamics of our brains be fundamentally quantum mechanical with the demand that our descriptions of our experiences of "what we have done and what we have learned" be essentially classical. This arrangement would be a natural upshot of the fact that our experiences would then correspond to the actualization of strictly quantum states that are both specified by classical states, and whose behaviors closely mimic the properties of their classical counterparts, *apart from the fact that they represent only potentialities*, and hence will be subject, just like Schroedinger' macroscopic cat, to the actions of the projection operators associated with our probing actions. This quantum aspect entails that, by virtue of the quantum Zeno effect, which follows from the basic quantum formula that connects our conceptually described observations to physically described quantum jumps, we can understand *dynamically* how our conscious choices can affect our subsequent thoughts and actions: we can rationally *explain*, by using the basic principles of orthodox contemporary physics, the causal efficacy of our conscious thoughts in the physical world, and thereby dissolve the physicalists' dilemma.

I shall now describe in more detail how this works.

## 8. The Mind-Brain Connection.

The general features of this quantum mechanical approach to the mind-brain problem have been described in several prior publications[9-12]. In this section I will present a specific model based on the general ideas described in those publications, but that adds some specifications pertaining to the quantum-classical connection..

Mounting empirical evidence[13,14] suggests that our conscious experiences are connected to brain states in which measurable components of the electromagnetic field located in spatially well separated parts of the brain are oscillating with the same frequency, and in phase synchronization. The model being proposed here assumes, accordingly, that the brain correlate of each conscious experience is an EM (electromagnetic) excitation of this kind. More specifically, each process 1 probing action is represented quantum mechanically in terms of a projection operator that is the quasi-classical counterpart of such an oscillating component of a classical EM field.

The central idea of this quantum approach to the mind-brain problem is that each process 1 intervention is the physical aspect of a psycho-physical event whose psychologically described aspect is the conscious experience of intending to do, or choosing to do, some physical or mental action. The physical aspect of the 'Yes' answer to this probing event is the actualization, by means of a quantum reduction event, of a pattern of brain activity called a "template for action". A *template for action* for some action X is a pattern of physical (brain) activity which if held in place for a sufficiently long time will tend to cause the action X to occur. The psycho-physical linkage between the felt conscious intent and the linked template for action is supposed to be established by trial and error learning.

A prerequisite for trial and error learning of this kind is that mental effort be causally efficacious in the physically described world. Only if conscious choices and efforts have consequences in the physically described world can an appropriate correlation connecting the mental and physical aspects of events be mechanically established by trial and error learning. With no such connection the physical action could become completely disconnected from the associated conscious intent with no adverse consequences.

The feature of quantum mechanics that allows a person's conscious choices to influence that person's physically described brain process in the needed way is the so-called "Quantum Zeno Effect". This quantum effect entails that if a sequence of very similar process 1 probing actions occur in sufficiently rapid succession then the affected component of the physical state will be forced, with high probability, to be, at the particular sequence of times $t_i$ at which the probing actions are made, exactly the sequence of states specified by the sequence of projection operators $P_h(t_i)$ that specify the 'Yes' outcomes of the sequence of process 1 actions. That is, the affected component of the brain state---for example some template for action---will be forced, with high probability, *to evolve in lock step with a sequence of 'Yes' outcomes of a sequence of "freely chosen" process 1 actions, where "freely chosen" means that these process 1 actions are not determined, via any known law, by the physically described state of the universe!* This coercion of a physically described aspect of a brain process to evolve in lock step with the 'Yes' answers to a sequence of process 1 probing actions *that are free of any known physically described coercion, but that seem to us to be freely chosen by our mental processes*, is what will presently be demonstrated. It allows physically un-coerced conscious choices to affect a physically described process that will, by virtue of the basic quantum probability formula, have intended experiential consequences.

In this model, the repetition rate (attention density) of the sequence of process 1 actions is assumed to be controlled by conscious effort. In particular, in the model where the projection operators $P(t_i)$ are projection operators $P[Q(t_i), P(t_i)]$---[In order to accommodate subscripts, I have now raised to *on-line* the arguments that in Chapter 7 I wrote as subscripts]---the presumption is that the size of the intervals $(t_{i+1} - t_i)$ are under the control of the psychological aspect of the probing action. This is in line with the general assumption that some of the details of the process 1 probing actions are at least partly under control of the associated stream of consciousness. This postulated influence of consciousness upon the *timings* of the events is the *only* influence granted to mind by this model. All other features of the events are assumed to be specified in some way by the physically described conditions.

I describe the quantum properties of the EM field in the formulation of relativistic quantum field theory developed by Tomonaga and by Schwinger. It generalizes the idea of the Schroedinger equation to the case of the electromagnetic field. One can imagine space to be cut up into very tiny regions, in each of which the values of the six numbers that define the electric and magnetic fields in that region are defined. In case the field in that region is executing simple harmonic oscillations we can imagine that each of the six values is moving in a potential well that produces the motion of a SHO. If the process 1 action is specified by a projection operator $P$ corresponding to a 'Yes' state that is a coordinated synchronous oscillation of the EM field in many regions, $\{R_1, R_2, R_3, \ldots\}$ then this state, if represented quantum mechanically, consists of some quasi-classical state $P[Q_1, P_1]$ in $R_1$, *and* some quasi-classical state $P[Q_2, P_2]$ in $R_2$, *and* some quasi-classical $P[Q_3, P_3]$ in $R_3$, etc.. The state $P$ of this combination is the *product* of these $P[Q_i, P_i]$s, each of which acts in its own SHO space, and acts like the unit operator (i.e., unity or 'one') in all the other spaces. This product of $P_n$ s, all evaluated at time $t_i$, is the $P_h(t_i)$ that is the brain aspect of the 'Yes' answer to the process 1 query that occurs at time $t_i$. The *quantum frequency* of the state represented by this $P_h(t_i)$ is the sum of the *quantum frequencies* of the individual regions, and is the total number of quanta in the full set of SHOs. However, the period of the periodic motion of the classical EM field remains $2\pi$, in the chosen units, independently of how many regions are involved, or how highly excited the states of the SHOs in the various regions become. This smaller frequency is the only one that the classical state knows about: it is the frequency that characterizes the features of brain dynamics observed in EEG and MEG measurements.

The sequence of $P_h(t_i)$s that is honed into observer/agent's structure by trial and error learning is a sequence of $P_h(t_i)$s that occurs when the SHO template for action is held in place (via the quantum Zeno effect) by effort. Learning is achieved by effort, which increases attention density, and holds the template for action in place during learning. Thus if $H_0$ is the Hamiltonian that maintains this SHO motion then for the honed sequence

$$P_h(t_{i+1}) = \exp\left(-\,iH_0\,(t_{i+1} - t_i)\right) P_h(t_i) \exp\left(iH_0(t_{i+1} - t_i)\right).$$

But in the application situation there may be disturbing physical influences that tend to cause a deviation from the learned SHO motion. Suppose that on the time scale of $(t_{i+1} - t_i)$ the disturbance is small, so that the perturbed evolution starting from $P_h(t_i)$ can be expressed in the form

$$P(t_{i+1}) = \exp\left(-iH_i\,(t_{i+1} - t_i)\right) \exp\left(-\,iH_0\,(t_{i+1} - t_i)\right) P_h(t_i)$$
$$\exp\left(iH_0(t_{i+1} - t_i)\right) \exp\left(\,iH_i\,(t_{i+1} - t_i)\right)$$

$$= \exp\left(-iH_i\,(t_{i+1} - t_i)\right)) P_h(t_{i+1}) \exp\left(\,iH_i\,(t_{i+1} - t_i)\right)$$

where $H_i$ is bounded.

According to the basic probability formula, the probability that this state $P(t_{i+1})$ will be found, if measured/observed, to be in the state $P_h(t_{i+1})$ at time $t_{i+1}$ is (using Trace $P_h(t_i) = 1$)

$$\text{Trace } P_h(t_{i+1}) \exp\left(-iH_i\,(t_{i+1} - t_i)\right)) P_h(t_{i+1}) \exp\left(\,iH_i\,(t_{i+1} - t_i)\right).$$

Inserting the leading and first order terms $[\,1\pm iH_i\,(t_{i+1} - t_i)]$ in the power series expansion of $\exp\left(\pm iH_i\,(t_{i+1} - t_i)\right)$ and using PP= P, and the fact that Trace AB = Trace BA, for all A and B, one finds that the term linear in $(t_{i+1} - t_i)$ vanishes identically.

The vanishing of the term linear in $(t_{i+1} - t_i)$ is the basis of the quantum Zeno effect. If one considers some finite time interval and divides it into small intervals $(t_{i+1} - t_i)$ and looks at a product of factors $(1 + c(t_{i+1} - t_i)^n)$, then if n is bigger than one the product will tend to unity (one) as the size of the intervals $(t_{i+1} - t_i)$ tend to zero. But this means that, if the initial answer is 'Yes', then the basic probability formula of quantum mechanics entails that, as the step sizes $(t_{i+1} - t_i)$ tend to zero, the evolving state of the system being probed by the sequence of probing action will have *a probability that tends to one (unity) to evolve in lock step with the set of 'Yes' answer, specified by the sequence of projection operators $P_h(t_i)$ associated with the previously learned action*. Thus the power merely to influence only the *intention densities* of a repetitious sequence of the process 1 actions confers upon conscious effort the power both to instill in the neuroplastic brain, by trial and error learning, physical templates for action associated with felt intentional effort, and, later, to hold in place by a similar mental effort the physical templates for the previously learned action. These "attention densities", though causally efficacious in the brain, are not themselves determined by any known law or rule. Hence they and their physical consequences *could be*, as they seem to us to be, influenced by conscious mental effort per se.

## 9. Conclusion.

Large-scale brain dynamics can be largely controlled by macroscopic brain activities that generate classically describable oscillating states of the electromagnetic field measured by EEG and MEG procedures. These states contain huge amounts of energy, on the atomic scale. Nevertheless, if we accept the principle that the underlying brain dynamics *must in principle* be treated quantum mechanically, then we must replace these classically conceived brain activities by their quasi-classical quantum counterparts. These are physically described SHO projection operators, each of which specifies the potentiality for the arising in the brain of a particular template for action. *The physical structure underlying each such template is instantiated in the neuroplastic brain, in association with a conscious effort, by a natural quantum process that exploits the quantum Zeno effect. A subsequent activation of this template can be sustained by a similar later conscious effort via the same quantum- Zeno-based process.* Thus the principles of orthodox (von Neumann/Heisenberg) quantum mechanics provide the dynamical basis for the natural creation causal mind-brain connections that allows an escape from the horns of the physicalists' dilemma. It gives each person's effortful conscious intentions the *direct* power to causally influence the course of events in his or her quantum mechanically described brain in a way that tends to produce the intended experiential feedback.

A further development and discussion of the mathematical details can be found in reference 15.

## Acknowledgement.

I thank Ed Kelly for useful suggestions pertaining to the form of this paper.